# Validating the Effectiveness of Data-Driven Gamification Recommendations: An Exploratory Study

Armando M. Toda[1,2], Paula T. Palomino[1], Luiz Rodrigues[1], Wilk Oliveira[1], Lei Shi[2], Seiji Isotani[1], Alexandra Cristea[2]

[1] Institute of Mathematics and Computer Science – University of São Paulo (ICMC-USP) São Carlos, SP – Brazil.

[2] Department of Computer Science – University of Durham – Durham, U.K

```
{armando.toda, paulatpalomino, lalrodrigues, wilk.oliveira}@usp.br,
          {lei.shi, alexandra.i.cristea}@durham.ac.uk
```

***Abstract.*** *Gamification design has benefited from data-driven approaches to creating strategies based on students' characteristics. However, these strategies need further validation to verify their effectiveness in e-learning environments. The exploratory study presented in this paper thus aims at verifying how data-driven gamified strategies are perceived by the students, i.e., the users of e-learning environments. In this study, we conducted a survey presenting 25 predefined strategies, based on a previous study, to students and analysed each strategy's perceived relevance, instanced in an e-learning environment. Our results show that students perceive Acknowledgement, Objective and Progression as important elements in a gamified e-learning environment. We also provide new insights about existing elements and design recommendations for domain specialists.*

## 1. Introduction

Gamification design[1] has become a topic of interest in recent studies in the field of computers in education. Many of these studies aim at providing guidelines on how to implement gamification properly, including in the education domain [Borges *et al*., 2014; Dichev & Dicheva, 2017]. However, literature states that gamification must be designed based on user characteristics, to ensure positive effects [Toda, Vale & Isotani, 2018; Santos, Bittencourt & Vassileva, 2018]. Prior works show also that gamification does not provide enough empirical evidence or enough data to attest its efficiency, which hinders the gamification adoption by domain specialists (teachers and instructors) [Toda *et al*., 2018b].

An alternative to tackling this issue is using the *data-driven approach* (*e.g.*, machine learning and data mining techniques) to analyse data provided by gamification/game-based systems, in order to provide empirical evidence on the gamification design [Meder, Plumbaum & Albayrak, 2017]. Although only a few studies have been conducted in this field, most of them are focused on the domain of business [Meder, Plumbaum & Albayrak, 2017]. Furthermore, these design strategies must be validated with users in order to provide their efficiency [Dichev, Dicheva and Irwin, 2018; Klock *et al*., 2018a].

---

[1] The process / method of creating gamified tasks

This work thus aims at providing a validation process of the findings of data-driven gamification recommendations proposed by us in Toda *et al* (2019), where the authors used a data mining approach to provide gamification strategies based on user gender. We conducted a mixed epistemological research, through an inductive method in a semi-controlled environment [Gomes & Gomes, 2019]. The objective of this paper is to explore and explain how the data-driven gamification strategies would be perceived by real users. To this end, we created the following research question "*How do users perceive game elements implemented in e-learning environments, based on prior research on data-driven gamification recommendations?*". The contributions of this work are aligned with the field of Gamification in Education, providing validated recommendations that could be used in the development and improvement of gamified educational systems. These contributions are:

- Approach to validating data-driven gamification recommendations;
- Gamification recommendations that can be used by the domain specialists when designing e-learning environments;
- New insights into how users' perceptions are influenced by the game elements implemented in the environment.

## 2. Related Works

In the work of Klock *et al*., (2018b), the authors demonstrated how learning analytics could be used within a gamified e-learning environment to provide a better insight for teachers and instructors. Although relevant and providing many insights into the field, it does not focus/addresses how these analytics could impact on the gamification design process (*e.g.*, demonstrating which elements are more preferable by certain users).

Shi and Cristea (2016) explored how to approach gamification based on the theoretical underpinning of the Self-Determination Theory (SDT). The authors provided a novel way of concretely implementing SDT-rooted game elements in e-learning environments towards increasing student motivation. Their work also used a survey to validate the proposed gamification recommendations. However, unlike our current study, their work only focused on gamifying and improving social interaction features in e-learning environments.

Another work, conducted by Pedro & Isotani (2016), focused on identifying and understanding students' performance and gender differences towards *"gaming the system"* (behavior associated with cheating in learning environments). Their results indicate gender differences towards preferences of game elements in the environment. The female population had an overall lower performance, compared to males, which may be associated to the competition-driven elements (*e.g.* points, leaderboards and levels). The authors did not focus on giving or analysing gamification recommendations, but they inferred that female individuals might not like competitive environments.

Finally, the work of Toda *et al.* (2019) used Association Rule Mining (ARM) to identify relations between user gender and game elements in a given dataset. They provided 25 rules converted into gamification strategies. However, the authors did not validate those rules with real users, but stated that other demographic variables (*e.g.*, age group) could be used to provide more focused rules.

In addition, for the related works above, current research in gamification is concerned in understanding how the combination of game elements influence on the perception of students and users, when using gamified systems [Gárcia Iruela et al., 2019; Featherstone and Habgood, 2019].

## 3. Methods and Tools

To conduct this study, we analysed previous data-driven gamification strategies proposed in Toda *et al.* (2019). Following that work, with the aid of specialists in the field of gamification and e-learning environments, we created 25 mockups (see an example mockup in Figure 1) on how these strategies would be applied in a generic e-learning environment. Then, we created a survey, where the users would state the relevance of that gamification strategy in the context it was applied (that of a generic e-learning system). The survey was validated with 3 gamification specialists, before being sent to the users. The strategies can be seen on https://forms.gle/4FdMfLsE3zbjo9Lu9.

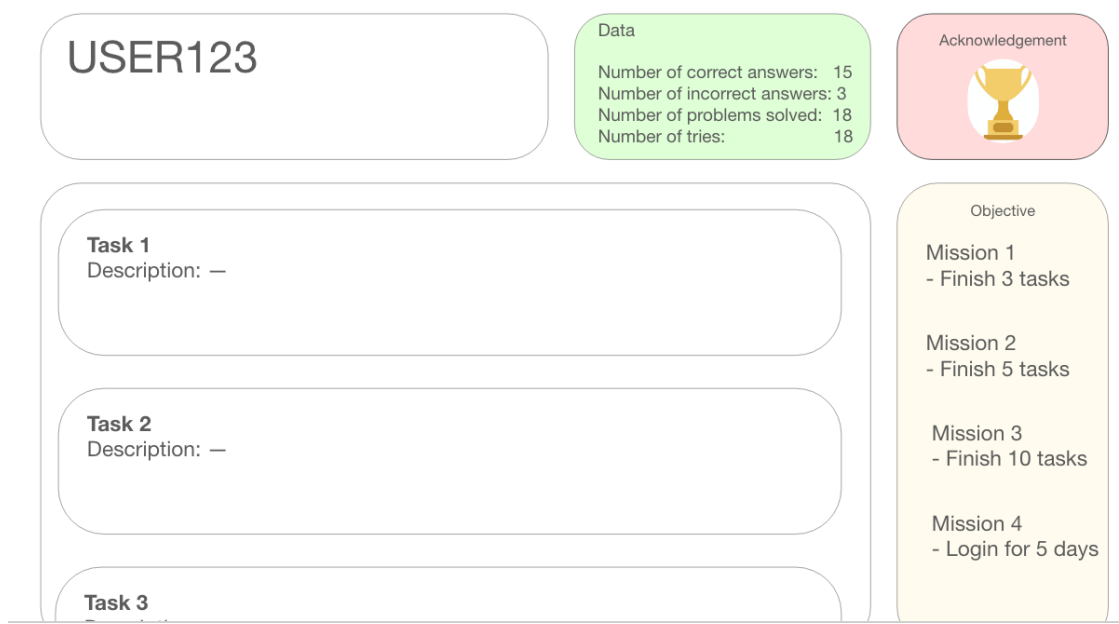

**Fig 1. Example of a Mockup (related to Table 1 below, Id 10) that was presented to the users.**

The relevance of the mockups was measured with non-polarised questions, in a non-biased way on a Likert scale [Likert, 1932] from 1 (Totally not relevant) to 5 (Totally relevant). Furthermore, each mockup was followed by a section where users could give open feedback and add comments, *e.g.*, "*I dislike Renovation*" or "*I really liked this layout*". We aimed at recruiting people with the same demographics - male and female, age group between 20 and 30 years, originally from Brazil (convenience sampling [Saunders, Lewis and Thornhill, 2009]). Henceforth, the recruitment process was conducted through email invitation. In total, we sent the survey to approximately 50 people.

All users would first need to agree that their data would be used for scientific purposes. The survey contained 29 questions: three questions were related to user demographics, one question on their perception of gender influence in game elements preference, and 25 questions directly about the mockups created. These mockup questions were divided into two groups, based on the gender (Male or Female). The users would receive the groups of questions related to their assigned gender first, then the opposite gender on the second group of questions. We made the survey this way, in order to explore how user gender might influence their perception of the data-driven gamification strategies. The users did not know about these randomised groups until the survey was submitted. The data used in this paper can be found at https://bit.ly/2ZlqwGM. After collecting the data, we analysed it through descriptive statistics (i.*e.*, Mean and Standard Deviation) in the general group and per gender. Next, we analysed the comments of each mockup and summarised the results, for positive (*e.g.* "*I liked this element*" or "*I think this combination might work well*") and negative comments (e.g. "*I disliked this element*" or "*I don't think this works*").

## 4. Results

In total, 15 out of 50 (30%) of the invited people answered the survey, which constitutes our focus group. The distribution is almost similar between male (N=8) and female (N=7) populations. The minimum age is 18 and the maximum is 38 (Mean = 28.2, SD = 5.8). Concerning the education level, five students are at postgraduate level, six are at least at the undergraduate level, two at high school level, and one at technical and one at elementary school level. A summary of this data is presented in Figure 2.

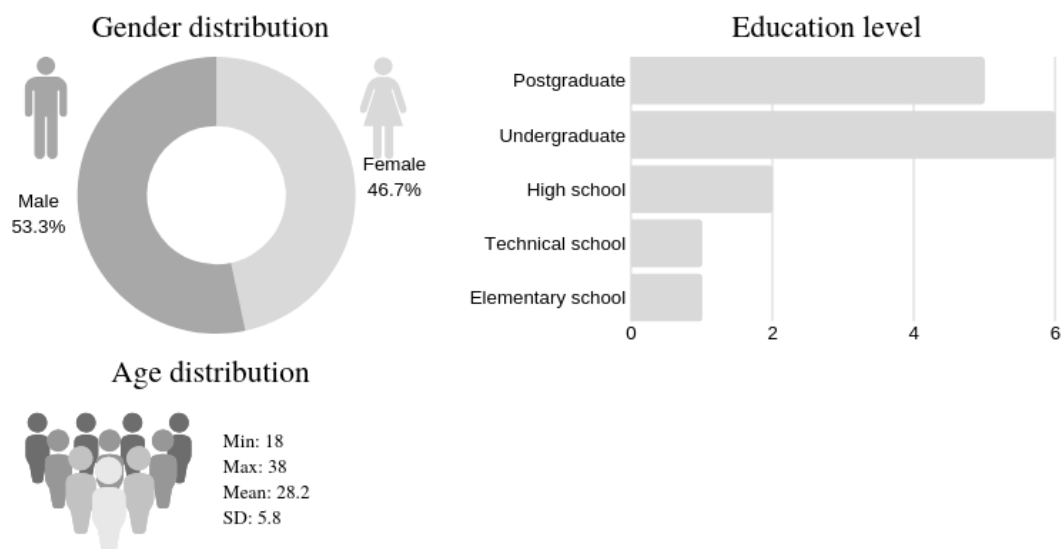

**Figure 2: Summary of the demographics**

Concerning the mockups, overall, the majority received a medium score from the users (respondents), with only one (Id 13, Mean = 2.7) achieving a score below the indifference point *i.e.*, 3, and two (Id 14 and 18, Mean = 4.2) above the relevance point *i.e.*, 4. A comparison of the overall results, and the results per genders can be seen in Table 1.

**Table 1. Overall and gender analyses**

| Id | Elements in the mockup | Overall | | Female | | Male | |
|----|------------------------|---------|-----|--------|-----|------|-----|
| | | Mean | SD | Mean | SD | Mean | SD |
| 1 | Progression-Data-Acknowledgement-Objective | 3,8 | 1,1 | 3,8 | 1,5 | 3,8 | 0,5 |
| 2 | Progression-Level-Acknowledgement-Objective | 3,6 | 1,5 | 3,6 | 1,9 | 3,7 | 1,2 |

| 3 | Progression-Acknowledgement-Objective | 3,8 | 1,1 | 3,3 | 1,2 | 4,3 | 0,8 |
| --- | --- | --- | --- | --- | --- | --- | --- |
| 4 | Level-Acknowledgement-Objective | 3,6 | 1,1 | 3,8 | 1,3 | 3,4 | 1 |
| 5 | Point-Acknowledgement-Objective | 3,8 | 1,1 | 3,9 | 1,3 | 3,7 | 1,1 |
| 6 | Progression-Puzzle-Objective | 3,6 | 1 | 3,8 | 0,8 | 3,4 | 1,2 |
| 7 | Novelty-Puzzle-Objective | 3 | 1 | 3,3 | 1,2 | 2,8 | 0,9 |
| 8 | Acknowledgement-Novelty-Objective | 3 | 1,1 | 3 | 1 | 3 | 1,2 |
| 9 | Acknowledgement-Choice-Objective | 3,4 | 0,9 | 3,2 | 0,9 | 3,5 | 0,8 |
| 10 | Data-Acknowledgement-Objective | 3,2 | 1,2 | 3,2 | 1,5 | 3,3 | 0,9 |
| 11 | Acknowledgement-Puzzle-Objective | 3,6 | 1 | 3,3 | 1 | 3,9 | 1 |
| 12 | Renovation-Progression-Choice-Objective | 3,7 | 0,9 | 3,8 | 1,2 | 3,7 | 0,8 |
| 13 | Progression-Social Pressure-Data-Objective | 2,7 | 1,2 | 2,5 | 1,3 | 2,9 | 1,2 |
| 14 | Progression-Puzzle-Acknowledgement-Objective | 4,2 | 0,9 | 3,8 | 1 | 4,5 | 0,6 |
| 15 | Level-Renovation-Progression-Objective | 3,5 | 1,1 | 3,5 | 1,2 | 3,5 | 1,1 |
| 16 | Point-Level-Puzzle-Objective | 3,2 | 1,1 | 4 | 0,6 | 2,5 | 0,8 |
| 17 | Progression-Level-Choice-Puzzle-Objective | 3,4 | 0,9 | 3,6 | 1 | 3,2 | 0,9 |
| 18 | Progression-Acknowledgement-Data-Objective | 4,2 | 0,9 | 4 | 1,2 | 4,3 | 0,5 |
| 19 | Progression-Level-Choice-Objective | 3,6 | 0,9 | 3,8 | 1 | 3,4 | 0,8 |
| 20 | Progression-Novelty-Data-Choice-Objective | 3,2 | 1,2 | 3,3 | 1,4 | 3,2 | 1 |
| 21 | Progression-Novelty-Acknowledgement-Choice-Objective | 3,4 | 1,2 | 3,2 | 1,3 | 3,5 | 1,1 |
| 22 | Progression-Level-Data-Choice-Objective | 3,4 | 1 | 3,5 | 1,2 | 3,3 | 0,9 |
| 23 | Progression-Novelty-Economy-Objective | 3,7 | 1 | 3,5 | 1 | 3,9 | 1 |
| 24 | Progression-Choice-Economy-Objective | 3,6 | 1,2 | 3,3 | 1 | 3,8 | 1,3 |
| 25 | Progression-Renovation-Novelty-Objective | 3,2 | 1,2 | 3,2 | 1,1 | 3,2 | 1,3 |

As we can observe in Table 1, the perceived relevance of these combinations is relatively similar for most mockups, for both males and females. We highlighted the lowest (red) and highest (blue) in each column to identify which was the most preferred by each gender, based on the "Overall" mean. Concerning the mockups, the least preferred mockup (Id 13), contains Social pressure, which is the use of direct or indirect peer interaction as a way of motivating and engaging the users/players [Dignan, 2011]. In our mockup, we used a space that presented updates on other users' activities, as a way of creating a motivational pressure from their peers. As we can observe, this may not be the best way to present this element, since users didn't like it, as well as complained about it in their comments. Concerning the highest scoring strategies, we can observe that both genders preferred Progression, Acknowledgement and Objective elements.

Progression, according to Toda *et al.* (2019) refers to the visualisation of progress of the users within an environment. In our mockups, this concept was implemented using a *progress bar* that could measure the number of activities that were made by the user (student). Acknowledgement is related to achievements/trophies, which are rewards given for specific actions within the environment, *e.g.*, 'the student finished 5 tasks in a row'. This was represented as trophies in our mockups. Objective is related to the learning goals, and was presented as 'missions'. These were original strategies found more relevant to males, in the work of Toda *et al.* (2019) and, in this work, were also found more relevant by the male individuals.

Concerning the comments, mockup 13 received the highest number of comments (9). Seven of these users states their "dislike" for the Social Pressure concept, stating that it may demotivate the final user by comparing them to other users. Only one person stated that they were pleased with the Social pressure concept. According to two comments, Social Pressure would be good for competitive users. Still concerning this mockup, the Data concept received two "dislikes" from the respondents. In our mockups, Data is presented as the visualisation of the user information as a number of correct and wrong tasks.

Mockups 12 and 23 received, respectively, a larger number of comments (8). Concerning mockup 12, most of the comments (5) stated that Choice was misplaced within this mockup, while four comments praised the use of Renovation. Choice was presented as a poll where the users (students) could choose their next assignment. Two comments stated that Choice should be something more meaningful, such as choosing the next subject. Renovation was presented through a re-do button, providing the student with the opportunity to perform the same task again. As for mockup 23, seven comments praised Economy while one said it did not like the element. Economy was presented in our mockup as virtual currency that could be spent in the system. Comments stated that Economy was an interesting element but the items that could be used as transactions should be more meaningful; there was also a statement about Economy being more relevant than Point. One comment also disliked Objective.

Mockups 2, 15 and 16 received seven comments each. For mockup 2, at least two users stated that Level and Progression should be one single element. In our mockups, Level was presented as a number that increased as the user gained points or completed a task. Two other comments stated that Level was indifferent to them and that the representation was rather confusing. One comment stated that they preferred Puzzle instead of Level in this mockup. On mockup 15, six comments stated they disliked Level; two suggested to group it with Progression. One comment disliked Renovation. As for mockup 16, five comments disliked the way the elements were presented while one comment stated that points should be used to buy things inside the system (change to Economy) and the other stated that Acknowledgement should be in place of Objective.

Mockups 1 and 5 received six comments each. On mockup 1, users stated that Data might not be that relevant, or should not be visible on the screen all the time. Other two comments implied that there were too many elements on the screen. One final comment suggested Data to be linked with Achievement. As for mockup 5, four comments stated they were displeased with Point, stating that this concept might be irrelevant when not being linked to Progression. Only one comment praised the use of Point in this mockup, stating that it should be great for continuous systems, where there was no progression towards the end.

Concerning mockups 4, 8, 20 and 24, each received five comments. For mockup 4, four users were displeased with Level, stating that it should be tied to Progression; one user further stated their preference of Acknowledgement over Level. One person disliked Objective. Regarding mockup 8, four users did not like the way Novelty was presented. Novelty in our work was presented as a space presenting updates on the system activities and changes. One user stated that Novelty should contain more relevant information, and others said that Novelty was distracting. One user that praised Novelty said that the screen contained too much information. Mockup 20 received overall mixed comments, where users stated that Novelty and Choice could be shown in other sections of the system. Data received similar comments as before, since two comments affirmed that it could be presented through Progression or linked with Acknowledgement. Finally, in this subset of comments, mockup 24 also received some mixed reviews, since users stated that Choice should have another type of presentation. Besides, two users were pleased with the Economy.

When reaching the threshold of four comments per mockup (4, 6, 7, 9, 10, 17 and 25), most of these comments repeated what previous ones stated. Mockup 4 received mixed comments, where two stated their preference towards this mockup while the other two made suggestions on how to adequately represent Acknowledgement. Mockup 6 also received mixed comments, one stated that Puzzle should be linked to a type of reward, while one user liked the concept of Puzzle. In this work, we represented Puzzle as extra challenges where users could improve their scores and test their skills. Mockup 7 received similar feedback, where two users stated they did not like Novelty yet the other two were pleased. As for mockup 9, three comments praised the combination suggesting that Choice should be presented in another way, while one disliked the concept. Mockup 10 was disliked by four comments: users stated that Data should be linked to other elements and Acknowledgement should be tied to Progression. In mockup 17, two comments suggested that Level and Progression should be one single element or tied to Acknowledgement. One user stated they were displeased with Choice while other stated their preference towards it. As for mockup 25, the comments also presented mixed opinions: one disliked Choice; one disliked Objective; and two praised the combination and Renovation element.

Finally, under four comments, we have mockups 11, 14, 18, 19, 21 and 22. Mockup 11 and 14 received three praises from the users, where one of them stated it was the best combination. It is worth noting that Mockup 14 also achieved a high score overall. Mockup 18 received the same previous comments towards Data (also achieved the highest score along mockup 14). Mockup 19 had overall mixed reviews: one comment stated disliking Choice while the others suggested a link between Level and Acknowledgement. Mockup 21 received two positive comments: one suggested to change the way Choice was presented, and the other suggested to remove Objective. Finally, mockup 22 had two comments towards Data, that repeated what was stated before.

## 5. Discussions

Based on our results, we did not find significant differences in terms of the perception of the mockups, between the two genders, possibly also due to this being only a focus study, with few users. Moreover, interestingly, we can observe from Table 1 that the mockups received an almost similar score from both genders. This may occur because of the small sample size and thus may need further in-depth investigation. We can also observe that, overall, all the mockups received a similar relevance score near the average point (3).

When analysing the comments, we can observe that most of them referred to changing the way an element was presented or linking it to Acknowledgement / Progression for the sense of progress and meaningfulness. These analyses also allow us to infer some statements towards this data:

- Users perceive Level and Progression as the same thing, although in the concept defined by previous works Toda *et al.* (2019), Level is usually related to the users' skills while Progression is related to the system status. Nevertheless, it is understandable why most of the users stated their preference to link those two elements, since most games and gamified environments also tie levels and progression (e.g. Duolingo).
- Progression, Objective and Acknowledgement are usually well accepted when linked together, rewarding the users with medals/trophies/achievements whenever they complete a task in the e-learning environment.
- Choice and Data should be treated as part of the system and have a separate visualisation from the gamified user interface. Choice would be meaningful if linked to Acknowledgement (e.g., allowing the student to win different trophies based on their personal choices in the system). Data should be presented in a more subtle way (e.g., discreetly scattered through the mockup) and provide relevant information.
- Point is not relevant, according to all people who commented; i.e., it should not be used alone. Point should be linked to other elements to be more meaningful. Similar remarks were made about Data.
- Acknowledgement is the most praised element, which can be used to reward students as well as to make other elements more meaningful to them (such as Level and Objective).

Our work confirmed the data-driven strategies from prior work: Progression, Objective and Acknowledgement as the main elements and most accepted by the users. However, we did not find significant differences towards the genders.

As for the limitations of our work, a major one is the sample size (15) which is very low. When we asked the participants why they did not answer the survey, most of them stated that it was either too confusing or too long to complete (the average time to complete was 20 minutes). Due to this sample size, we cannot confirm using statistics if our results are significant. However, as it is known from the Human Computer Interaction literature that five users could be an acceptable sample to conduct a usability study[2], we believe that even though not achieving statistical significance, our results may be meaningful and able to provide insight from the users' perspective for designers and other domain specialists in the field.

## 6. Conclusions and Future Work

This work presented a validation of gamification recommendations for e-learning environments using a data-driven approach. This validation was conducted based on the students' perception of the game elements in mockups predefined based on prior research outcomes clustering gamification elements perceived as working well together. Through this study, we observed that the combination of elements do influence students' perception of the system. In this work we have shown faint differences between males and females in terms of their perception of gamified elements and their pairing, such as female individuals preferring Progression element more than male individuals, while the second prefer Acknowledgement more than female individuals, which is the opposite from prior findings. However, we have found that Progression, Acknowledgement and Objective may be a suitable combination for both genders.

---

[2] https://www.nngroup.com/articles/how-many-test-users/

As future work, we intend to conduct a large-scale validation (with a larger sample size) on the relevance/acceptance of the combinations of the gamification concepts and game elements, in order to verify statistical differences towards their relevance to different genders. We also will be focusing on embedding these recommendations in a recommender system, to support the gamification design.

## Acknowledgements

The authors would like to thank the funding provided by FAPESP (Projects 2018/15917-0; 2016/02765-2; 2018/11180-3; 2018/07688-1), CAPES and CNPq.